\documentclass[twocolumn]{aastex631}

\shorttitle{L dwarfs detection from SDSS images using improved
Faster R-CNN} \shortauthors{Zhi Cao et al.}

\graphicspath{{./}{figures/}}
\usepackage{graphicx}   
\usepackage{amsmath}    

\usepackage{amssymb}    
\usepackage{multirow}
\usepackage{supertabular,booktabs}
\usepackage{threeparttable}
\usepackage{makecell}
\usepackage{array}
\usepackage{bm}

\begin{document}

\title{L dwarfs detection from SDSS images using improved Faster R-CNN}

\author{Zhi Cao}
\affiliation{School of Mechanical, Electrical and Information Engineering, Shandong University, 180 Wenhua Xilu,  Weihai, 264209, Shandong, China\\}

\author{Zhenping Yi} 
\altaffiliation{yizhenping@sdu.edu.cn}
\affiliation{School of Mechanical, Electrical and Information Engineering, Shandong University, 180 Wenhua Xilu,  Weihai, 264209, Shandong, China\\} 
\author{Jingchang Pan} 
\affiliation{School of Mechanical, Electrical and Information Engineering, Shandong University, 180 Wenhua Xilu,  Weihai, 264209, Shandong, China\\}
\author{Hao Su} 
\affiliation{School of Mechanical, Electrical and Information Engineering, Shandong University, 180 Wenhua Xilu, Weihai, 264209, Shandong, China\\} 
\author{Yude Bu}
\affiliation{School of Mathematics and Statistics, Shandong University, 180 Wenhua Xilu, Weihai, 264209, Shandong, China}
\author{Xiao Kong} 
\affiliation{CAS Key Laboratory of Optical Astronomy, National Astronomical Observatories, Beijing 100101, China} 
\author{Ali Luo} \affiliation{CAS Key Laboratory of Optical Astronomy, National Astronomical Observatories, Beijing 100101, China} 

\begin{abstract} 
We present a data-driven approach to automatically detect L dwarfs from Sloan Digital Sky Survey (SDSS) images using an improved Faster R-CNN framework based on deep learning. The established L dwarf automatic detection (LDAD) model distinguishes L dwarfs from other celestial objects and backgrounds in SDSS field images by learning the features of 387 SDSS images containing L dwarfs. Applying the LDAD model to the SDSS images containing 93 labeled L dwarfs in the test set, we successfully detected 83 known L dwarfs with a recall rate of 89.25\% for known L dwarfs. Several techniques are implemented in the LDAD model to improve its detection performance for L dwarfs, including the deep residual network and the feature pyramid network. As a result, the LDAD model outperforms the model of the original Faster R-CNN, whose recall rate of known L dwarfs is 80.65\% for the same test set. The LDAD model was applied to detect L dwarfs from a larger validation set including 843 labeled L dwarfs, resulting in a recall rate of 94.42\% for known L dwarfs. The newly identified candidates include L dwarfs, late M and T dwarfs, which were estimated from color ($i-z$) and spectral type relation. The contamination rates for the test candidates and validation candidates are 8.60\% and 9.27\%, respectively.
 The detection results indicate that our model is effective to search for L dwarfs from astronomical images. 

\end{abstract}




\section{Introduction}

Ultracool dwarfs are stellar or sub-stellar objects include extremely low-mass stars at the bottom of the main sequence and brown dwarfs \citep{BurrowsHubbard-8,2002ApJ...567..304C},  with spectral types M7V or later, including late M, L, T and Y \citep{10.1093/mnras/staa3423}. Ultracool dwarfs are the lowest-mass, coldest and faintest products of star formation  with temperature no higher than 2700\,K \citep{kirkpatrick1995solar}. They represent about 15\% of the astronomical objects in the stellar neighborhood of the Sun \citep{cantrell2013solar}. Since their minimal core fusion mostly preserves their natal composition, ultracool dwarfs are yardsticks of Galactic chemical evolution \citep{2019ApJ...883..205B}, and provide a unique probe of large-scale Galactic structure and evolution \citep{Aganze_2022}.

Ultracool dwarfs are faint and radiate most light in the red optical and near-infrared bands. Searching for point-like objects meeting certain color criteria in optical and near-infrared band images is an effective way to find ultracool dwarf candidates.  For instance, the optical color $(i-z)$ was found to be sensitive to ultracool dwarfs \citep{Fan2000,Hawley_2002} , and the $(i-z)$ cut has been applied to help find ultracool dwarfs \citep{Chiu_2006,Zhang2009,SchmidtWest-12}. Late L dwarfs and T dwarfs radiate more strongly in near-infrared band, and thus the near-infrared colors $J-H$, $H-K_s$, $Y-J$ have been used to select ultracool dwarf candidates \citep{ Kirkpatrick1999,Burgasser2002,Cruz2003,2016A&A...589A..49S, RosellSantiago-19}.
 More than ten thousand ultracool dwarfs have been obtained through selection using photometric colors, combined with spectroscopy and/or proper motion.\citep{2019ApJ...883..205B}.

Large-area surveys in optical and near-infrared bands, such as the Sloan Digital Sky Survey (SDSS; \citealt{2000AJ....120.1579Y}), the Two Micron All Sky Survey (2MASS; \citealt{SkrutskieCutri-31}), and the UKIRT Infrared Deep Sky Survey (UKIDSS; \citealt{LawrenceWarren-33}), have contributed significantly to the discovery of ultracool dwarf candidates.
The next generation of imaging surveys, e.g., Large Synoptic Survey Telescope, (LSST; \citealt{2009arXiv0912.0201L}), Chinese Space Station Telescope (CSST; \citealt{2011Consideration}), will yield large amounts of high-quality data. Maximizing the extraction of useful and reliable information from the huge amount of data is one of the goals pursued by the next generation of sky surveys \citep{Ivezic2016}. Machine learning offers a promising approach for extracting important features and maximizing the use of astronomical data. In previous studies, several machine learning algorithms have been employed to perform classification, parameters measurement and celestial object search. For example, a convolutional neural network (CNN) model was built to classify stars and galaxies \citep{10.1093/mnras/stw2672}. \citet{10.1093/mnras/sty3217} built a Bayesian neural network to measure multi-element abundances in stellar spectra. 
\citet{2021AJ....162..155X} applied a CNN model to stellar spectra to search for very metal-poor stars. An object detection model was built to automatically detect low surface brightness galaxies from SDSS images \citep{10.1093/mnras/stac775}.  These new techniques automate the analysis and help to obtain more accurate results.

In this study, we aim to develop an L dwarf automatic detection (or LDAD for short) model to search for L dwarfs from SDSS field images using deep learning. Here, we choose to focus on L dwarfs which are among the faint ultracool dwarfs, and there are enough resolved L dwarfs in SDSS, which can provide labeled samples for training a detection model utilizing a deep neural network. Object detection techniques are used to intelligently identify and locate L dwarfs in SDSS field images by learning the features in the training data without separating CCD data sources and extracting photometric parameters first. 
In contrast to common classification and regression tasks in astronomical applications, the task of this study includes not only the classification task of distinguishing L dwarfs from other celestial objects and backgrounds in astronomical images, but also the prediction of obtaining the coordinates of each L dwarf. Moreover, multiple L dwarfs can be identified simultaneously from the whole astronomical image, unlike the usual classification tasks where one image corresponds to one object. 

In this work, the deep neural network learned the features of L dwarfs in SDSS images to obtain the ability to identify L dwarfs. Compared to the $(i-z)$ cut typically used in selecting SDSS ultracool dwarfs, a well-designed deep neural network can maximize the extraction of features of L dwarfs in images, including colors, brightness and shape, etc. Using deep learning, features are automatically extracted and are combined in a complex nonlinear manner to optimise the detection. In addition, images of multiple bands can be easily used together to improve the detection accuracy. In this work, in addition to the most effective $i$ and $z$ bands, the $r$ band is used to provide additional information to help distinguish L dwarfs and exclude other type sources. The detection using our model is independent of SDSS photometry and some L dwarfs with unclean photometry could be detected. Therefore, the detection results using LDAD can complement color cut results. 

This paper outlines the method of detecting L dwarfs from SDSS images. In Section \ref{section:data}, we describe the sample selection of L dwarfs and image processing, as well as the preparation of training set and test set. The structure of the LDAD model is introduced in Section \ref{section:Model}. The performance of the LDAD model for detecting L dwarfs from the test set is shown in Section \ref{sec:Results}. In Section \ref{sec:Verification}, we use a larger L dwarf sample of SDSS to validate the model performance. Finally, conclusions are drawn in Section \ref{sec:CONCLUSIONS}.

\section{DATA} 
\label{section:data} 
To build a data-driven model, a reliable and labeled standard set of L dwarfs is essential. The object detection model LDAD takes a whole observed image containing L dwarfs as the input and outputs the coordinate position of L dwarfs in the image and the size of the bounding box, which is a rectangle surrounding an L dwarf. This section introduces the selection of L dwarf samples, the images preprocessing, and preparation of the training set and test set.

\begin{table}
\centering
    \caption{Reference list of 480 L dwarfs in our dataset}
    \label{tab:list}

    \begin{tabular}{lll} 
        \hline
        ID&source &number \\
        \hline
            1 & \citet{SchmidtWest-12}&116\\
    2 & \citet{2010MNRAS.404.1817Z}&103\\
     3 & \citet{2008AJ....135..785W}&88\\
    4 & \citet{2014AJ....147...34S}&39\\
    5 & \citet{2014ApJ...794..143B}&33\\
    6 &\citet{Kirkpatrick_2000}&20\\
     7 & \citet{2002AJ....123.3409H}&16\\
    8 & \citet{2009AA...494..949S}&9\\
     9 & \citet{Kirkpatrick1999}&5\\
     10 & \citet{Chiu_2006}& 4\\
    11 &\citet{2017AJ....153..196S}&4\\

    12 & \citet{2016ApJ...830..144R}&4\\
   13 & \citet{2004AJ....127.3553K} &3\\
      14 & \citet{refId0}&3\\

    15 & \citet{2017AJ....154..112K}&2\\
   16 & \citet{2015MNRAS.449.3651M}&2\\
      17 & \citet{2002ApJ...564..466G}&2\\
       18 & \citet{2002AJ....123..458S}&2\\
    19 & \citet{2003AJ....126.2421C}&2\\
      20 & \citet{2011ApJS..197...19K}&2\\
    21 & \citet{2014ApJ...792..119D}&2\\
   22 & \citet{2012ApJS..201...19D}&1\\
    23 & \citet{2015ApJ...814..118B}&1\\
     24 & \citet{2013PASP..125..809T}&1\\
    25 & \citet{2017MNRAS.464.3040Z}&1\\
    26 &\citet{2016ApJS..224...36K}&1\\
    27 &\citet{2015AJ....150..182K}&1\\
     28 &  \citet{2009ApJS..182..543A}&1\\
    29 &\citet{2013MNRAS.434.1005Z}&1\\
    30 & \citet{2009AJ....137..304S}&1\\
     31 & \citet{2016ApJ...828L..22S}&1\\
    32 &  \citet{2015ApJS..219...33G}&1\\
     33 &  \citet{2002ApJ...575..484G}&1\\
    34 &  \citet{2010ApJS..190..100K}&1\\
     35 &  \citet{2010ApJ...715..561A}&1\\
    36 &  \citet{2011ApJ...739...48A}&1\\
    37 &  \citet{Dupuy_2017}&1\\
    38 &  \citet{2017AJ....153...92T}&1\\
    39 &  \citet{2012AA...542A.105L}&1\\
    40 &  \citet{2017MNRAS.468..261Z}&1\\

        \hline
    \end{tabular}

\end{table}

\subsection{Data selection}
\label{section:Data selection}

The L dwarfs and images in this work were selected from the SDSS. The SDSS is a multi-band photometric and spectroscopic survey. SDSS images were taken using a photometric system of five filters ($u, g, r, i$ and $z$). To date, SDSS has encompassed more than one-third of the entire celestial sphere and presents a great opportunity for the study of various types of astronomical objects. In this work, the images containing L dwarfs were selected from SDSS DR16. We selected L dwarfs by cross-matching the sources of spectral type L in the SIMBAD database \citep{2000AAS..143....9W} with sources of SDSS, then removing the objects without clean photometry in $i, z$ bands  according to SDSS 'clean' flags\footnote{https://www.sdss.org/dr16/algorithms/photo\_flags\_recommend/}, resulting in 480 L dwarfs from 40 references, covering spectral types L0-L9. These references and the number of L dwarfs are shown in Table \ref{tab:list}. Then the SDSS field images containing these L dwarfs were obtained, with a total 480 images in $u, g, r, i, z$ band, which have been sky-subtracted and calibrated.

\subsection{Data preprocessing} 
\label{section:Data preprocessing} 
L dwarfs are faint and small objects, which emit radiation mostly in the far red of the optical and near-infrared band. Therefore, for SDSS data, we used the synthetic images of $r, i$ and $z$ bands to search for L dwarfs. The steps to synthesize $riz$ images are as follows: 
\begin{enumerate} 
\item Read in $r, i, z$ band fits files of SDSS images. 
\item Correct pixel shifts. There is a time difference between observations in each band, so the pixel position of the $r, i, z$ band needs to be corrected. We use Python's reprojection toolkit \citep{2018AJ....156..123A} to align the $i$-band and $z$-band to the $r$-band for correction according to the World Coordinate System (WCS) coordinate.
\item To scale $r, i, z$ bands properly, we apply arcsinh mapping according to Equation~(\ref{eq:generateImg1}), Equation~(\ref{eq:generateImg2}) and Equation~(\ref{eq:generateImg3}) \citep{Lupton_2004}. $i_{min}$, $r_{min}$ and $z_{min}$ are the minimum in the $i$, $r$ and $z$ band, respectively. 
\begin{gather}
        i_{1}=\frac{arcsinh(\alpha Q(i - i_{min}))}{Q} \label{eq:generateImg1} \\
        r_{1}=\frac{arcsinh(\alpha Q(r - r_{min}))}{Q}  \label{eq:generateImg2}\\
        z_{1}=\frac{arcsinh(\alpha Q(z - z_{min}))}{Q}
     \label{eq:generateImg3}
\end{gather} 
where $\alpha$ is the linear stretch parameter and $Q$ is the $asinh$ softening parameter. In this study, $\alpha=0.5$ and $Q=8$. \item synthesize RGB images using $i_{1}$, $r_{1}$ and $z_{1}$ , corresponding to the three primary colors red, green and blue, respectively. \end{enumerate}

\begin{figure}
\centering
    \includegraphics[width=\columnwidth]{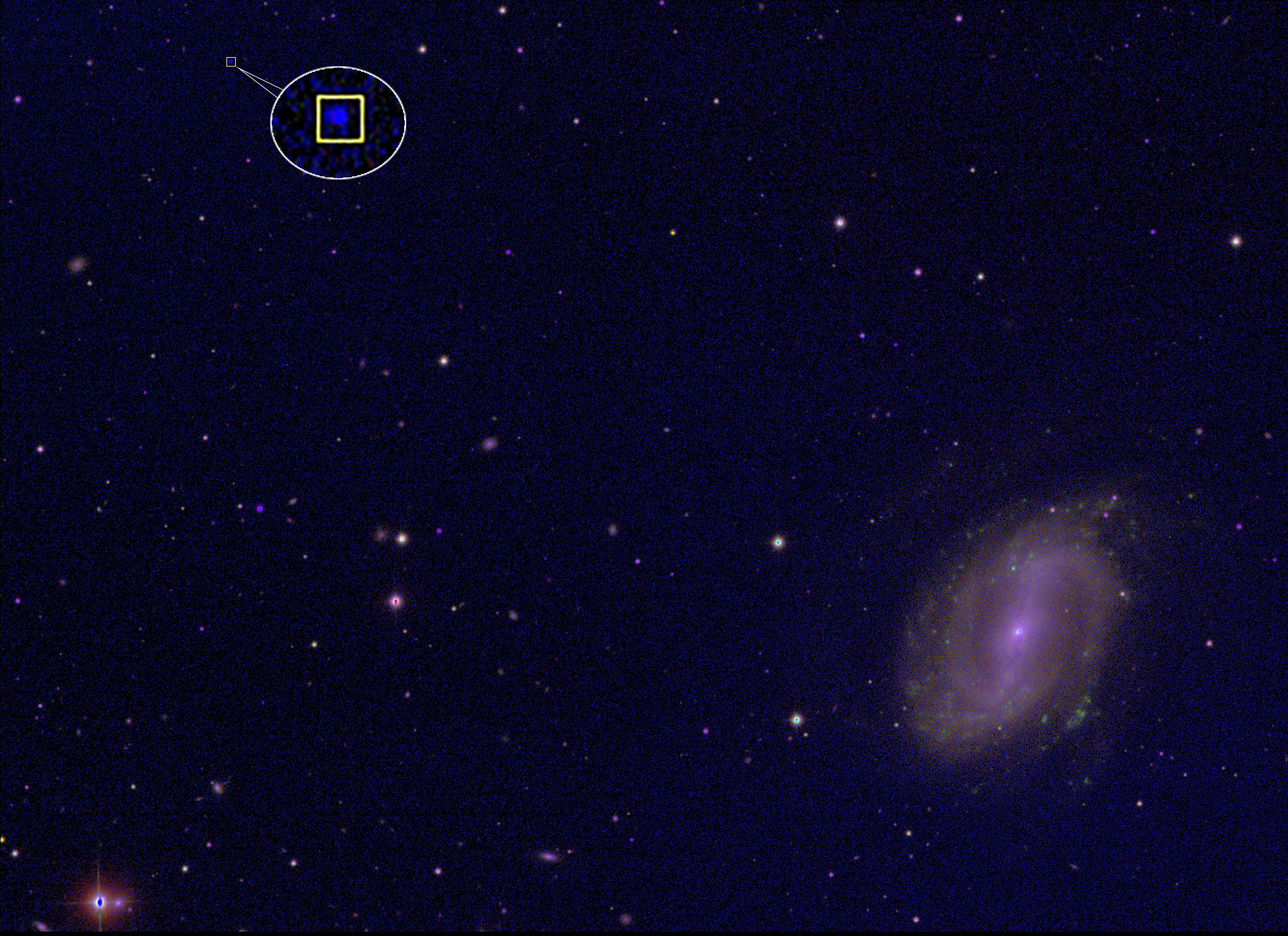}
\caption{An $riz$ synthesized image of SDSS with an L dwarf in a bounding box. The image is 1489 $\times$ 2048 pixels, while the bounding box in green is 15 $\times$ 15 pixels. In this field image, the L dwarf is so small in size that we can hardly  see it. In order to clearly see the L dwarf, the bounding box containing the L dwarf is enlarged by 5 times and shown in the white circle.}
    \label{fig:1}
\end{figure}

In addition to the SDSS images containing L dwarfs as the input to the target detection network, we need to prepare labels as the output of the network. The labels include the coordinate positions of the L dwarf in the SDSS image, and a bounding box of the appropriate size to frame the L dwarf. The R.A. and DEC. coordinates of L dwarfs were converted into the coordinates of pixels in the image. According to our statistics, the size of L dwarfs in our sample ranges from 5 $\times$ 5 pixels to 15 $\times$ 15 pixels. We
used bounding boxes with a fixed size of 15 $\times$ 15 pixels for all L dwarfs. For example, a processed SDSS image with an L dwarf in a bounding box of 15 $\times$ 15 pixels is displayed in Figure \ref{fig:1}.

\subsection{Training set and test set} 
\begin{figure}
\centering
    \includegraphics[width=\columnwidth]{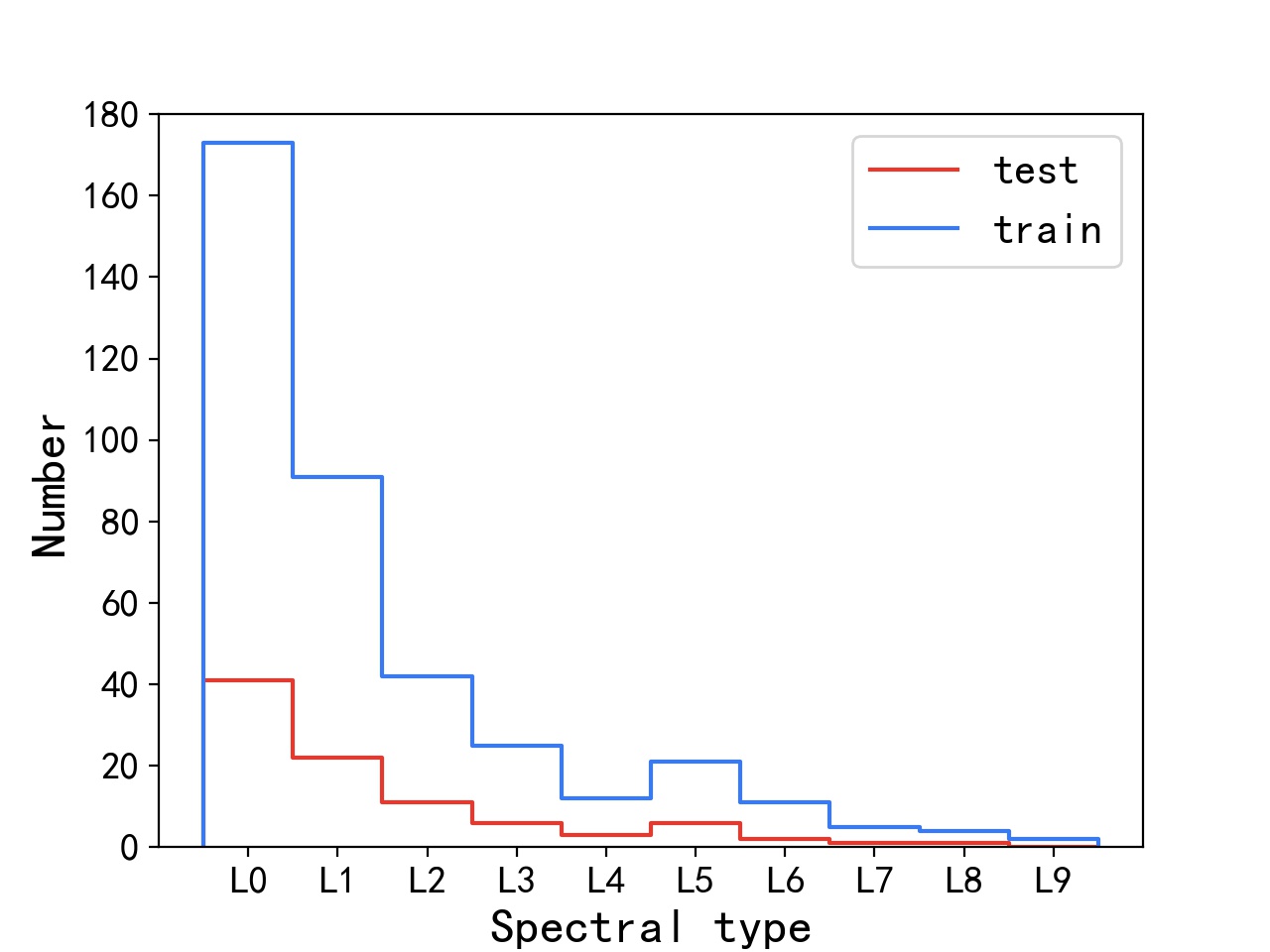}
\caption{The spectral type distribution of L dwarfs in the training set and test set.}
    \label{fig:distribution}
\end{figure}

Finally, we divided the processed images into a training set and a test set in a ratio of about 8:2. The training set includes 387 L dwarfs, while the test set includes 93 L dwarfs. The spectral type distribution of the training set and test set is depicted in Figure \ref{fig:distribution}. For the training set and test set, the spectral type distributions are similar. There are more early L dwarfs than late L dwarfs in both data sets. 

\section{Building the LDAD model} \label{section:Model} As an object detection model based on a deep neural network, the LDAD model incorporated some deep learning techniques, including residual neural network (Resnet: \citealt{He_2016_CVPR}), feature pyramid networks (FPN: \citealt{Lin_2017_CVPR}) for object detection and Faster R-CNN (\citealt{RenHe-59}). In this section, we introduce the techniques used in the LDAD network, and describe the structure, the loss function of the LDAD model and the training strategy.

\subsection{The object detection techniques used in LDAD}
\label{sec:technology}
Resnet is a special artificial neural network proposed by \citet{He_2016_CVPR} for alleviating the degradation problem of training very deep networks. In our model, the Resnet module is made up from residual blocks with bottlenecks. Each block has 1 $\times$ 1, 3 $\times$ 3 and 1 $\times$ 1 convolutional layers.  Each convolutional layer is followed by a rectified linear unit (ReLU) function, which is a commonly used activation function in image detection tasks to increase the sparsity of the model and facilitate the extraction of features. The input of bottleneck skips three convolution operations and adds directly to the result of the final ReLU activation function. This Resnet framework makes network optimization simpler and can obtain higher precision by increasing network depth.
In terms of size of L dwarfs, the searching targets are smaller than 15 $\times$ 15 pixels in a 1489 $\times$ 2048 pixel image. The smallest L dwarf is only 5 $\times$ 5 pixels, while the largest L dwarf is close to 15 $\times$ 15 pixels. The size of the largest L dwarf is nine times the size of the smallest L dwarf. Considering these features of L dwarfs in images, we added the structure of FPN for object detection to the model to extract features at multiple scales.

\begin{figure}[h]
\centering
    \includegraphics[width=\columnwidth]{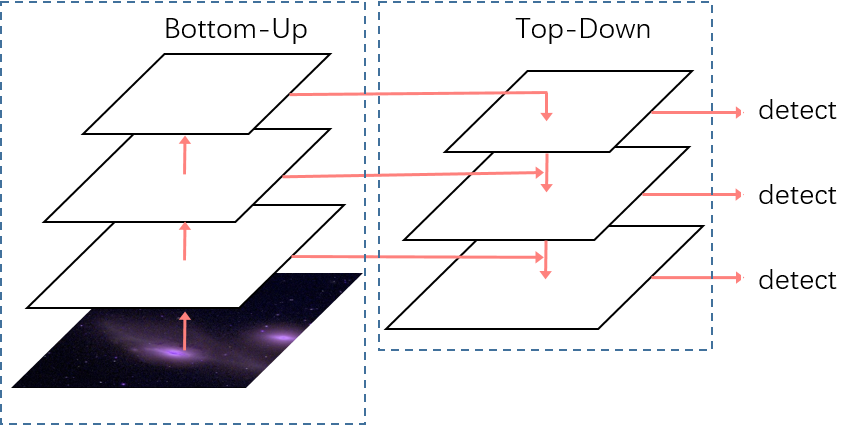}
\caption{Pyramidal hierarchy of deep convolutional networks to construct feature pyramids.}
    \label{fig:fpn}
\end{figure}

FPN is the inherent multi-scale, pyramidal hierarchy of deep convolutional networks to construct feature pyramids. As illustrated in Figure \ref{fig:fpn}, there are two main components: 
\begin{enumerate} 
\item Bottom-Up architecture: Calculate features at different scales and make features abstracted. 
\item Top-Down architecture with lateral connection: Starting from the most abstract high-level features and continuously integrating low-level
features. \end{enumerate}

The LDAD is developed based on the framework of Faster regions with convolutional neural networks (Faster R-CNN), which is one of the best object detection frameworks and has been widely applied in computer vision tasks. A Faster R-CNN object detection network consists of a feature extraction network, a region proposal network (RPN), a region of interest (ROI) pooling layer and a classification and localization network, as illustrated in Figure \ref{fig:f4}. The functions of each part are as follows:

\begin{figure}[h]
\centering
    \includegraphics[height=0.7\columnwidth]{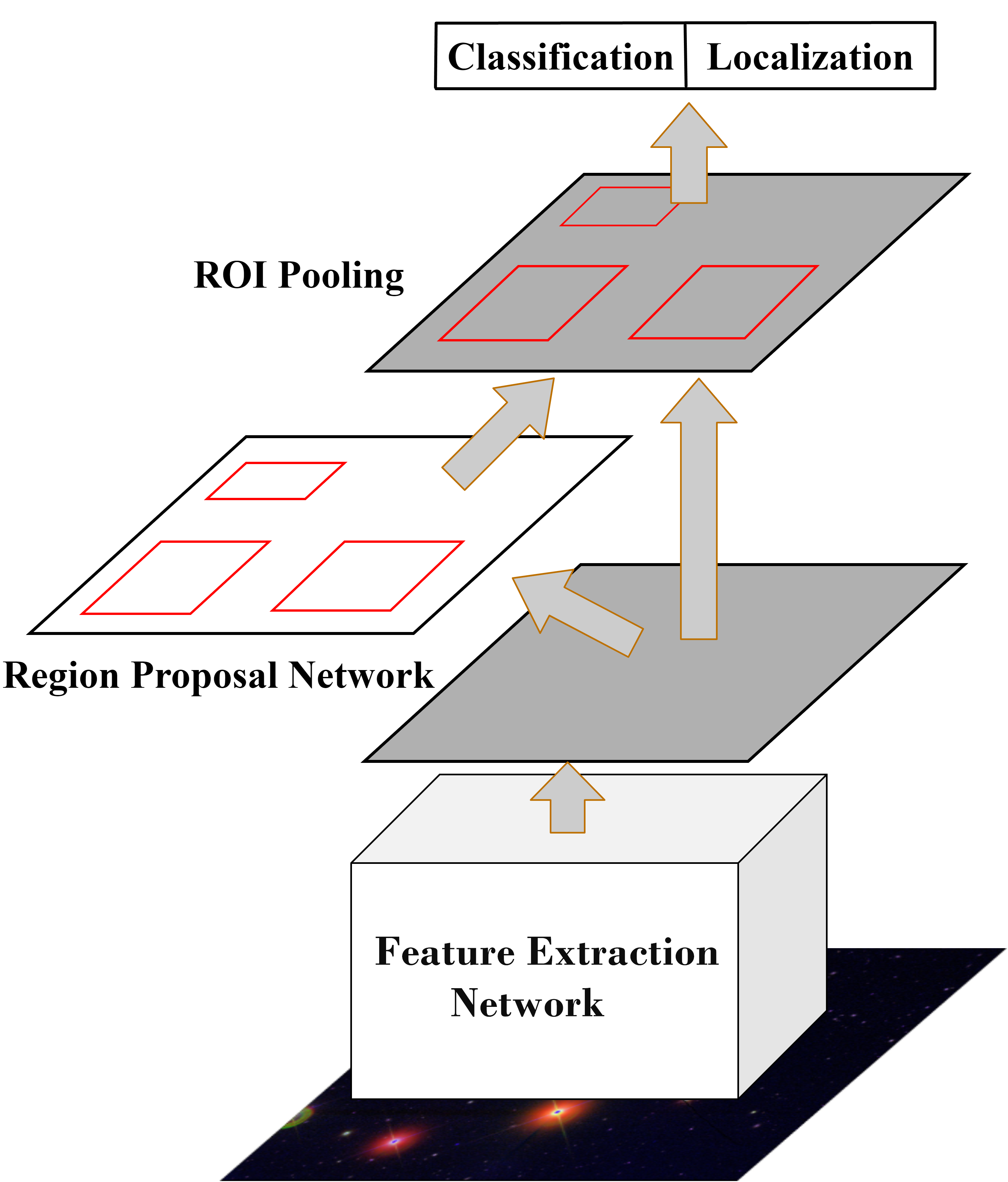}
    \caption{The framework of Faster R-CNN.}
    \label{fig:f4}
\end{figure}

\begin{enumerate} 
\item The feature extraction network uses a combination of basic convolutional layers, pooling layers and activation functions to extract feature maps for subsequent network use. \item The RPN finds out the regions where an L dwarf may exist in the image and generates a rough selection and position of the regions. 
\item ROI pooling transforms regions of different sizes that may have L dwarfs provided by the RPN network into features of the same size. 
\item The classification module gives the probability of the existence of L dwarfs in the candidate regions, and the localization module predicts the positions of L dwarfs in the candidate regions. 
\end{enumerate}

\subsection{The structure of LDAD model} 
\label{sec:structure} 
\begin{figure}
\centering
    \includegraphics[width=0.7\columnwidth]{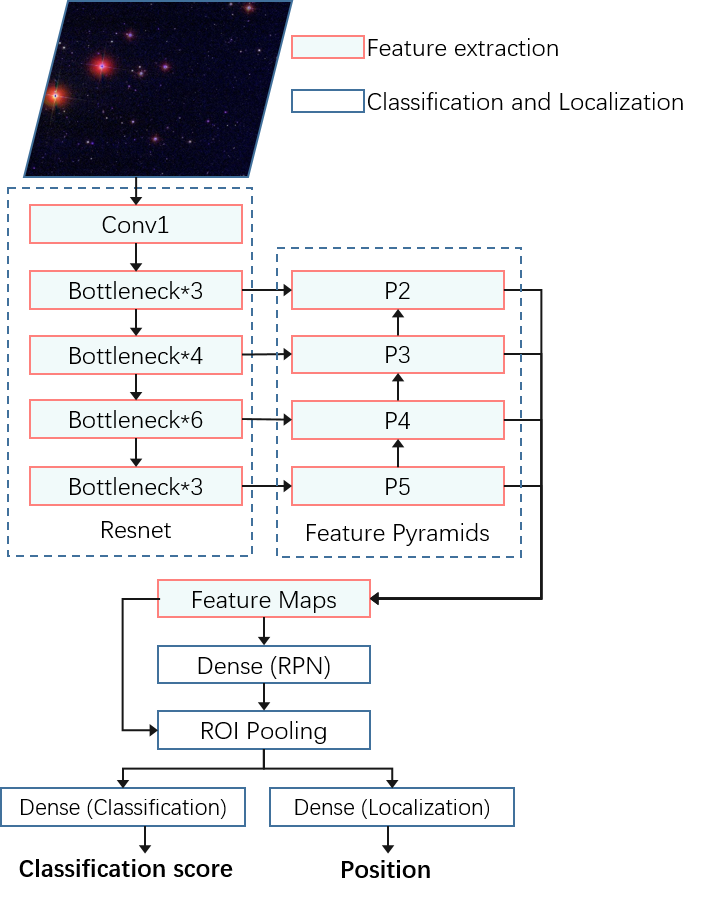}
\caption{The structure of the LDAD network. P2, P3, P4 and P5 are multi-scale features. The model is divided into two stages: find out candidate regions that may contain L dwarfs, and identify and locate L dwarfs in the candidate region. The input of the model is a 1489
$\times$ 2048 $riz$ image and the output is the positions and corresponding probabilities of all possible L dwarfs contained in the image. }
    \label{fig:f5}
\end{figure}

Our LDAD model has the same structure as the original Faster R-CNN, except that Resnet and FPN are used to replace the ordinary convolution network to achieve multi-level feature extraction. Its structure is depicted in Figure~\ref{fig:f5}. The workflow of the model is as follows:

\begin{enumerate} 
\item Image Processing. Conv1 contains a 7 $\times$ 7 convolution kernel with a stride of 2 and a 3 $\times$ 3 maximum pooling with a stride of 2. The image is processed in the conv1 module and the size of the image is reduced to 1/4 the original size while retaining the features of the original image as much as possible. Conv1 effectively reduces the amount of
computation of the model. 
\item Multi-level fusion feature extraction. Multiple bottleneck structures can extract the features of four levels of the image and fuse these features using the P2,
P3, P4 and P5 modules. 
\item Find the Region Proposals. The predefined anchor boxes, a set of predefined bounding boxes of a certain height and width, are tiled across the fusion feature map. The dense layer (fully connected network) in RPN predicts the probability of containing an L dwarf for every tiled anchor box. For the anchor boxes with the probability of containing L dwarfs greater than 0, the ROI pooling module divides them into fixed-size parts, and performs maximum pooling on each part to obtain region proposals of the same size. 
\item Predict classification score and position. The classification scores and L dwarf coordinates are predicted by a fully connected network. 
\end{enumerate}

\subsection{Loss function}
\label{sec:Loss}

A loss function is defined to measure the prediction ability of a machine learning model, which reflects the difference between the predicted value of the model and its actual value.  Training a model aims to reduce the loss gradually by a learning algorithm. Therefore, the definition of the loss function is crucial to the model. The loss function we used in the LDAD model is shown in Equation~(\ref{eq:loss}). The loss of the model consists of classification loss ($L_{cls}$) and regression loss ($L_{reg}$). As Equation~(\ref{eq:loss_cls}) affirms, the classification loss relies on the classic binary cross-entropy loss. The regression loss adopts Smooth L1 loss as expressed in Equation~(\ref{eq:loss_reg}).
\begin{gather}
     L(p_{i},t_{i})=\frac{1}{N_{cls}}\sum\limits_{i=1}^{}L_{cls}(p_{i},p^{*}_{i})+\frac{1}{N_{cls}}\sum\limits_{i=1}^{}p^{*}_{i}L_{reg}(t_{i},t^{*}_{i}) \label{eq:loss}\\
        L_{cls}(p_{i},p^{*}_{i})=-\log[p_{i}*p^{*}_{i}+(1-p_{i})(1-p^{*}_{i})] \label{eq:loss_cls}\\
         L_{reg}(t_{i},t^{*}_{i})=R(t_{i}-t^{*}_{i}) \label{eq:loss_reg}\\
         R(x)=\begin{cases}
         \frac{0.5x^{2}}{\beta}\quad \quad \quad \quad|x|<\beta\\
     |x|-0.5\beta \quad otherwise
     \label{eq:L1}
     \end{cases}
\end{gather} 
$i$ is the anchors index. $p_{i}$ is the classification score given by the model if it detects the object as an L dwarf. $p^{*}_{i}$ is the true classification of the object. If the area detected by the model contains an L dwarf, the value of $p^{*}_{i}$ is set to 1, otherwise it is set to 0. $t_{i}$ is the parametric coordinate of the L dwarf detected by the model, and $t^{*}_{i}$ is the parametric coordinate of the L dwarf in our training set in this area. $N_{cls}$ is the number of anchors. $\beta$ is a parameter that controls where the quadratic function and linear function are switched in Equation~(\ref{eq:L1}). In this work, it was set to 1/9 according to the implement of the faster R-CNN \citep{RenHe-59}.
\subsection{Training process}
\label{sec:traing}

In the training set, each SDSS image contains a labeled L dwarf. However, there may be other unknown L dwarfs in the image, which may be regarded as background or other types of celestial objects, playing the role of label noise, which may reduce the model's ability to identify L dwarfs. In order to avoid the negative impact of unknown L dwarfs on the performance of the model, we added an L dwarf noise removal module before the training.  We obtained the $i$-band and $z$-band magnitude of all objects in these images from the SDSS catalog and corrected them utilizing the extinction provided by SDSS. Then, we used color ($i-z$) to estimate the spectral type of each object in the images of the training set. (The color limit between L and M dwarfs is referenced from \citet{2015AJ....149..158S} while the color limit between L and T dwarfs is referenced from \citet{2015AA...574A..78S}.) If the estimated spectral type of an object is L and the object is not in our training set, the pixel values in its bounding box were set to 0, so as to eliminate the impact of unlabeled L dwarfs as label noises on the performance of the model. Through the noise removal, 781 objects were removed from the images in the training set.

The model was trained by applying classical gradient descent. As the loss was gradually reduced, the output value of the model gradually became close to the real value. The direction of loss descent is usually determined by the gradient of the loss function, while the learning rate determines how big a step is taken in that direction. The learning rate is a tuning parameter in an optimization algorithm that determines the step size at each iteration while moving toward a minimum of a loss function. We used an interval adjustment algorithm (stepLR) to adjust the learning rate dynamically with an initial learning rate of 0.005. During the training of the model, the learning rate continues to decrease. After 20 epochs of training, the loss function converged and the training process was completed.

\section{RESULT}
\label{sec:Results}
\subsection{Evaluation method}

We evaluate the LDAD model with recall ($R$), position deviation $\eta$,  precision ($P$) and contamination ($C$). Recall is the percentage of known L dwarfs that are detected by the model. $\eta$ is the average distance between the objects detected by the model and the objects in SDSS. Precision and contamination are measures of quality of detected samples.  These evaluations are calculated as follows:
  \begin{itemize}
      \item recall($R$):
\begin{equation}
        R=\dfrac{TP}{K}
\end{equation}
       \item position deviation $ \eta $:
       \begin{equation}
       \eta=\dfrac{1}{N} \sum_{n=1}^N distance(x_i,x_{SDSS})
       \end{equation}
         \item precision ($P$):
       \begin{equation}
       P=\dfrac{TP+L_{estimated}}{N}
       \end{equation}
        \item contamination ($C$):
       \begin{equation}
       C=1-P
       \end{equation}
  \end{itemize}
True Positive ($TP$) is the number of known L dwarfs detected by each model, $K$ is the number of known L dwarfs in the test set, $N$ is the number of objects detected by the model, $x_i$ is the R.A. and DEC. of the L dwarf candidates predicted by the model, and $x_{SDSS}$ is the R.A. and DEC. of SDSS L dwarfs in the training dataset. $L_{estimated}$ is the number of candidates of spectral type L which were estimated according to the relationship of color ($i-z$) and spectral type \citep{2015AJ....149..158S,2015AA...574A..78S}.

$p$ is the classification score of the object detected by the LDAD model, which takes values from 0 to 1; the larger the $p$ value, the more likely the identified object is an L dwarf. By setting the threshold of $p$, only the detected objects with score higher than the threshold are retained. As the threshold of $p$ is set to a larger value, the precision increases and contamination decreases accordingly, while the recall decreases. There is an inverse relationship between precision and recall. Therefore, we use a balanced measure f1-score, which is the harmonic mean of precision and recall. The f1-score is defined as follows: 
\begin{equation}
        f1\text{-}score=\dfrac{2*P*R}{P+R}
\end{equation}

\subsection{Detection results}
\label{sec:result}
\begin{table*}[!htb] 
\caption{Detection results of the models for the test set} 
\label{tab:Model test results} \centering
   \begin{threeparttable}
\begin{tabular}{*9{c}}
\hline
Model      & $N$\tnote{1} & $TP$\tnote{2}  & $FN$\tnote{3} & Candidate  &  $\eta$\tnote{4} & Recall&Precision&Contamination \\
 \hline
Faster R-CNN  &112 & 75                       & 18                                 & 37                       & $3.41^{\prime \prime}$                                  & 80.65\% & 69.64\% & 30.36\% \\
LDAD      &93 & 83                        & 10                                  & 10                         & $0.87^{\prime \prime} $                               & 89.25\% & 91.40\% & 8.60\%\\
\hline
\end{tabular}


 \begin{tablenotes}
       \footnotesize
       \item[1] $N$ is the number of objects detected by the LDAD model.
       \item[2] $TP$ is the number of known L dwarfs detected by the LDAD model.
        \item[3] $FN$ is the number of undetected known L dwarfs.
       \item[4] $\eta$ is the average distance error.
     \end{tablenotes}
   \end{threeparttable}
\end{table*}

We built the LDAD model by learning 387 training images. Then we use the model to detect L dwarfs from 93 $riz$ images in the test set. Running an Nvidia 2080ti GPU, it takes 11.03 seconds to detect 93 SDSS $riz$ images, and the average time to detect an $riz$ image is 0.1186 seconds.

\begin{figure}[!h]
\centering
    \includegraphics[width=\columnwidth]{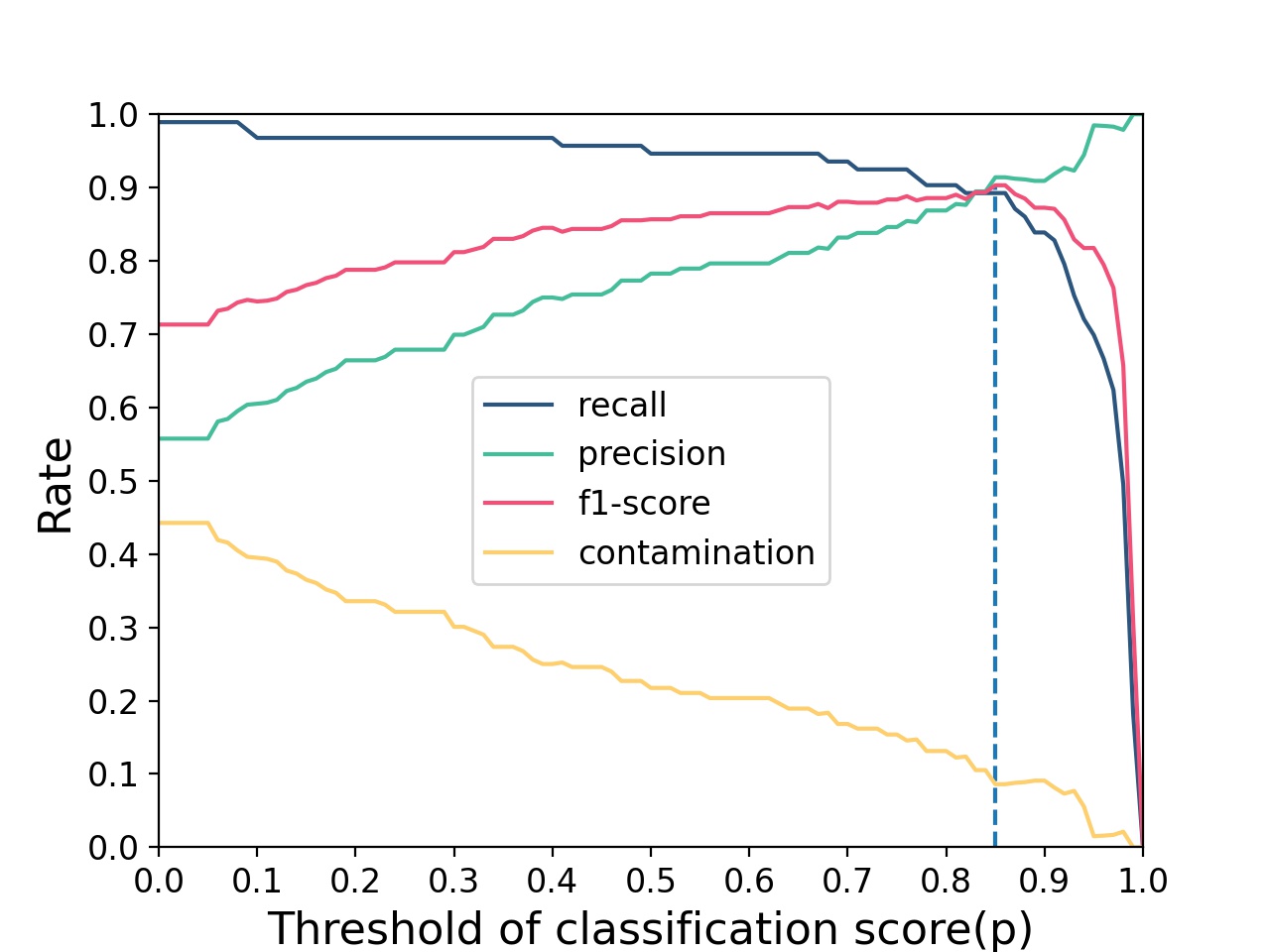}
\caption{Performance of recall, precision, f1\text{-}score and contamination under different score thresholds.}
    \label{fig:pr_img}
\end{figure}

The performance of recall, precision, f1\text{-}score and contamination under different score thresholds is shown in Figure~\ref{fig:pr_img}. With the increase of score threshold, precision and f1\text{-}score gradually increase, and recall slowly decreases until the threshold reaches 0.85. After the threshold value exceeds 0.85, recall and f1\text{-}score start to drop sharply. To achieve a better balanced performance between recall and precision, we set the score threshold of the model to 0.85.

\begin{figure}[h]
\centering
    \includegraphics[width=\columnwidth]{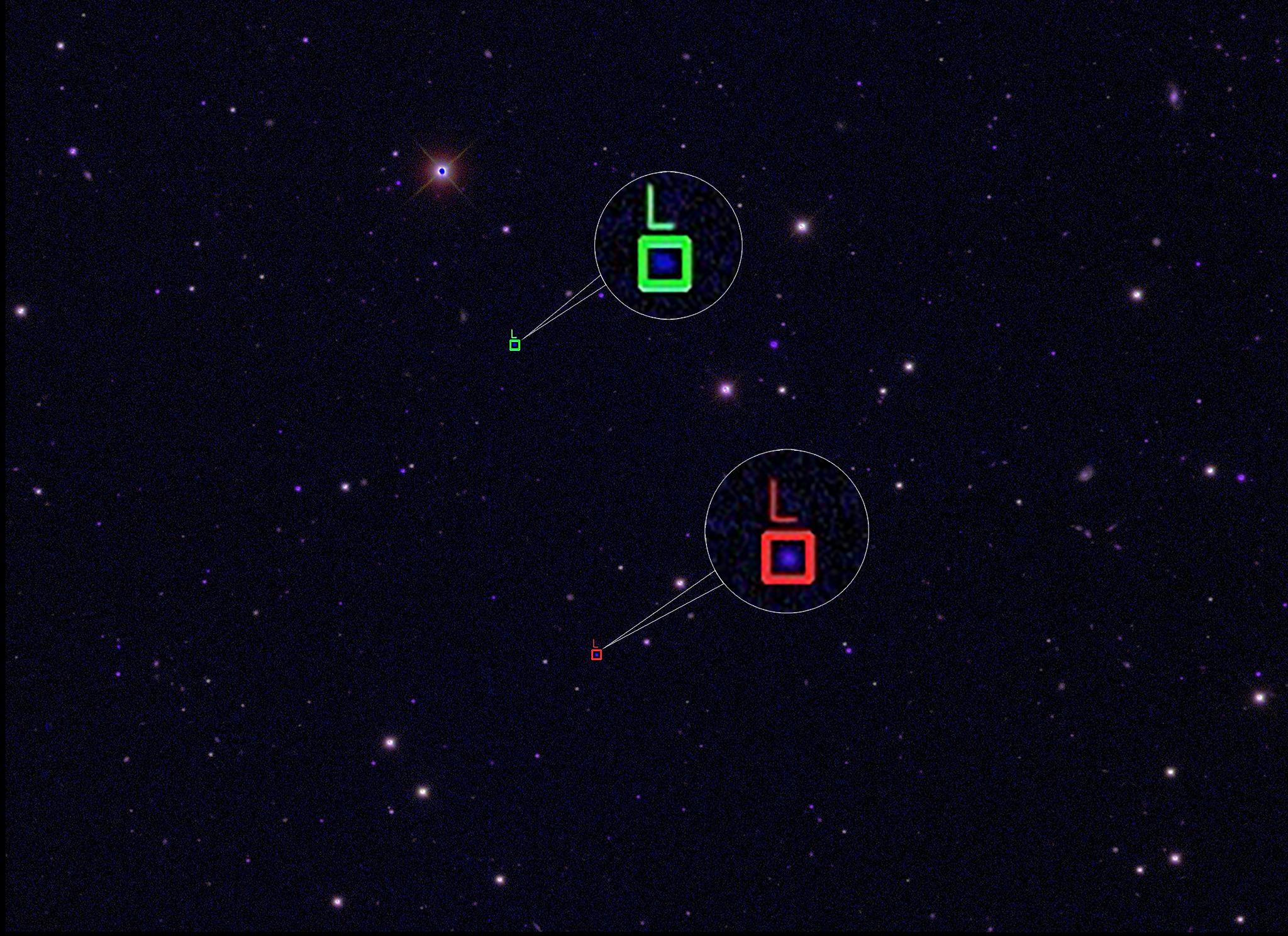}
\caption{Two objects were detected in an SDSS image of the test set using the LDAD model, including a known L dwarf in the green bounding box and an L dwarf candidate in the red one. The enlarged images of the corresponding original bounding boxes at the lower left are displayed in circles.}
    \label{fig:res_img}
\end{figure}

From 93 SDSS images in the test set, the model detected 93 objects, including 83 known L dwarfs and 10 candidates. The recall of our LDAD model is 89.25\%, while the  precision is  91.40\%. The average error between the coordinates of the detected L dwarfs and those in the SDSS catalog is 0.87 arcsec. (A more detailed description about the detection is given in Section 4.3.) Figure~\ref{fig:res_img} shows the two L dwarfs detected from an SDSS image. In this image, in addition to recalling a labeled L dwarf (in green bounding box), the model also detected a new L dwarf candidate (in red bounding box).

As a comparison, we trained a Faster R-CNN model according to \citet{RenHe-59} using our training dataset and detected L dwarfs from our test set. The detection results showed that recall and precision of the Faster R-CNN model are 80.65\% and 69.64\%, respectively. LDAD model outperformed the built Faster R-CNN model, with higher recall rate and precision, and correspondingly lower contamination. The detection results of the two models are displayed in Table \ref{tab:Model test results}.

\subsection{Result analysis}
\label{sec:Result analysis}


For the 10 newly detected L dwarf candidates, their spectral types were not found from SDSS or SIMBAD. We estimated their approximate spectral types using the color ($i-z$). As a result, they were classified into two L dwarfs, three M dwarfs and five T dwarfs. We listed their detailed information in Table \ref{tab:candidates}, in which $p$ is the classification score provided by our model. We noticed that some of the candidates do not have clean photometry, thus, the estimated spectral types may be unreliable. 

\begin{figure}
\centering
    \includegraphics[width=0.45\textwidth]{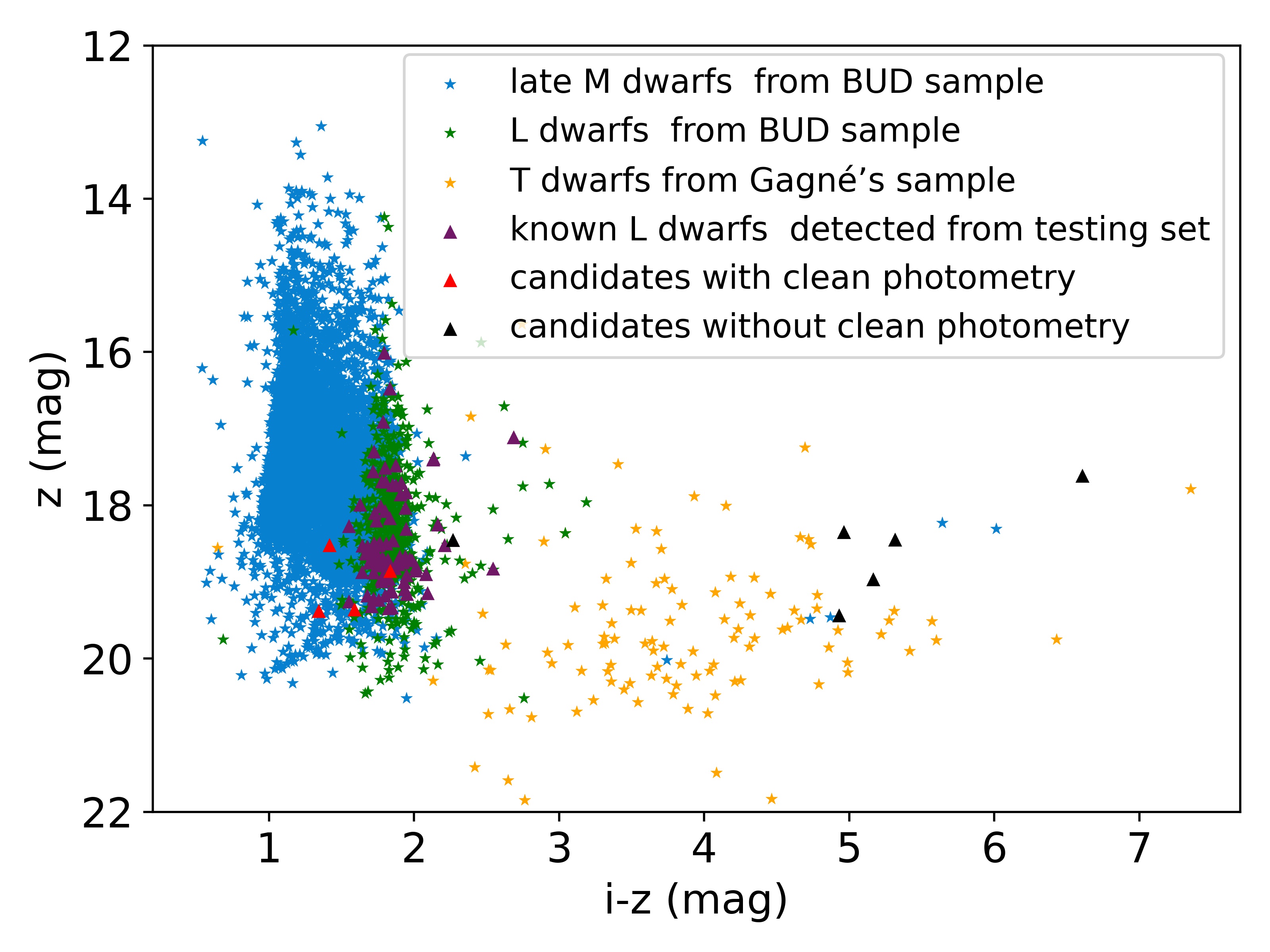}
\caption{Color-magnitude diagram of L dwarfs and L candidates detected by the LDAD model from the testing set. Three samples of late M dwarfs, L dwarfs and T dwarfs are added to the image for comparison.}
    \label{fig:i-z}
\end{figure}

The types of all detected objects from test images are summarized in Table \ref{tab:LDAD result}. Figure \ref{fig:i-z} shows the color and magnitude distribution of L dwarfs and candidates detected from the test set, as represented by triangles. The purple triangles represent the known L dwarfs detected from the test set, the red triangles correspond to detected candidates with clean photometry while the black ones signify candidates without clean photometry. For comparison, late M dwarfs and L dwarfs from the BUD sample \citep{2015AJ....149..158S} are marked in Figure \ref{fig:i-z} as blue and green asterisks, respectively, and a T dwarf sample including 117 T dwarfs obtained by crossing (the cross radius is 3 arcsec) Gagn\' e's sample\footnote{https://jgagneastro.com/list-of-ultracool-dwarfs/} with SDSS is displayed as orange pentagrams in Figure \ref{fig:i-z}. Overall, the color ($i-z$) distribution of the objects detected by the LDAD model is consistent with that of L dwarfs in the BUD sample except for five with unreliable photometry. This indicates that the L dwarfs detected by the LDAD model are generally reliable.

\begin{table*}[] \caption{The objects newly detected by the LDAD model} \label{tab:candidates} \centering \scriptsize
\begin{tabular}{llllllll} \hline
$objID$(SDSS)&R.A.         & DEC.        &clean flag ($i$)&clean flag ($z$)& $i-z$ (extinction corrected)& type&$p$ \\
\hline
1237657856066585449&    157.4825551&    48.46527762&    clean& clean&1.4161 &$M_{estimated}$&   0.9418\\
1237651250441618107&    245.416571  &45.70010626    & clean& clean& 1.5904& $M_{estimated}$&    0.9409\\
1237662663753139329 &238.618466&    27.24267& clean& clean& 1.3436& $M_{estimated}$ &0.9086\\
1237659133282353373&    216.9811153&    51.37158449 & unclean& clean&   2.2668& $L_{estimated}$&    0.9454\\
1237662474234955244&    251.0952676&    25.97624751 & clean& clean& 1.8356& $L_{estimated}$&    0.9340 \\

1237657235452330196&    43.98702888&    1.113097469 &unclean&unclean&   6.6060  &$T_{estimated}$&   0.9829\\
1237648722290279036&    147.5770404&    0.751733445 &unclean&unclean&   4.9279  &$T_{estimated}$&   0.9412\\
1237651821637338187&    219.9115324&    3.354177331 &unclean&unclean&   4.9613  &$T_{estimated}$&   0.9369\\
1237667550346150696&    167.3698958&    23.75943152 &unclean&unclean&   5.3158  &$T_{estimated}$&   0.9312\\
1237662499470509079&    241.9480368&    35.03123429 &unclean&unclean&   5.1637  &$T_{estimated}$ &  0.9154\\

\hline
\end{tabular}
\end{table*}

\begin{table}[h]\scriptsize
\centering
\caption{The spectral types of objects detected by the LDAD model from the test set}
\label{tab:LDAD result}
\begin{tabular}{ccccc}
\hline
Class      & L  &  $M_{estimated}$& $L_{estimated}$  & $T_{estimated}$   \\
 \hline
Number     & 83     &  3                         & 2        & 5       \\
Proportion & 89.25\% &  3.23\%    & 2.15\%&  5.37\%    \\
\hline
\end{tabular}
\end{table}

\begin{figure*}
    \includegraphics[width=\textwidth ]{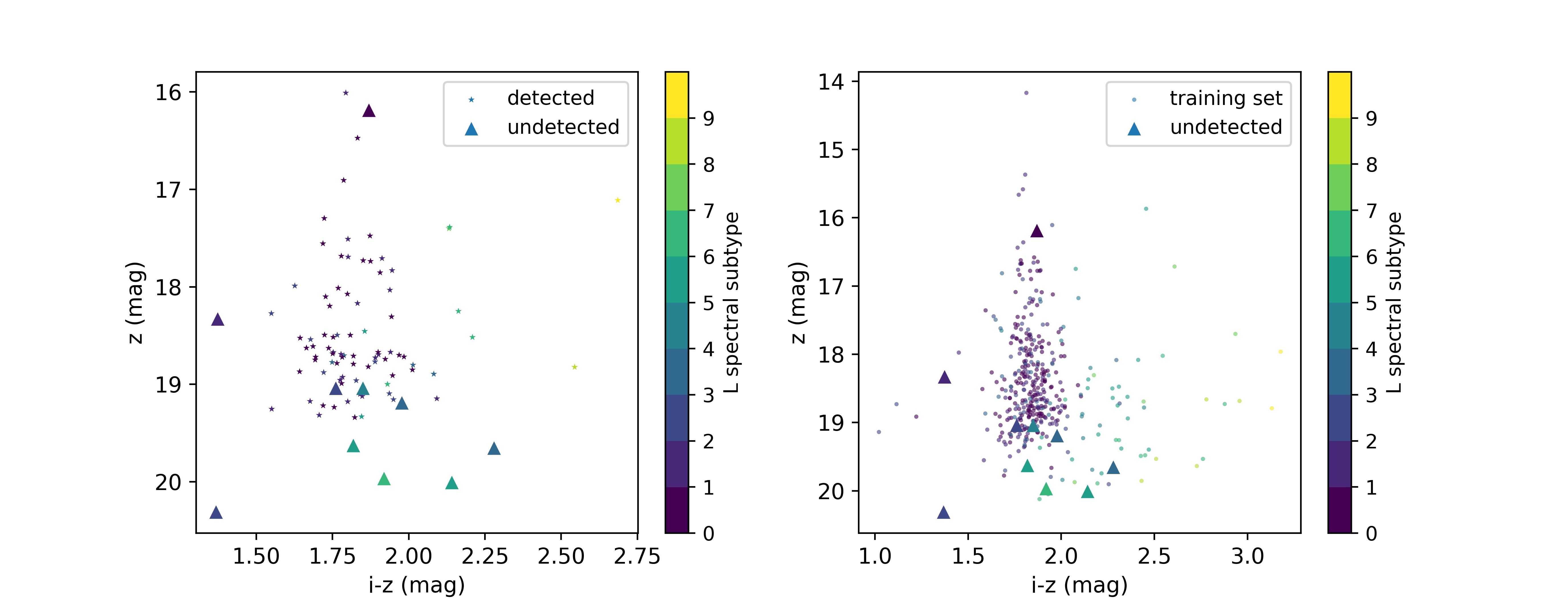}
\caption{The spectral subtype distribution with color ($i-z$) and $z$-band magnitude of detected and undetected L dwarfs by the LDAD model. Asterisks and triangles represent the L dwarfs that are detected and undetected from the test dataset, respectively, while circles signify the L dwarfs in the training dataset. Objects are color-coded by spectral subtype of L dwarf.}
    \label{fig:f9}
\end{figure*}

Ten labeled L dwarfs were not detected from the test image by the LDAD model. Their spectral subtype distribution with color ($i-z$) and $z$-band  magnitude is depicted in Figure \ref{fig:f9}. The left panel shows the detected (asterisks) and undetected (triangles) L dwarfs from the test images, while the right panel features the undetected L dwarfs from the test dataset (triangles), and L dwarfs in the training dataset (circles) as a comparison. In Figure \ref{fig:f9}, objects are color-coded by spectral type.
In this figure, we find that undetected L dwarfs are either very faint ($z$\textgreater19 mag) or at the outer boundary of the training samples. There are few training samples in these areas, resulting in insufficient feature learning for the samples in these regions by the model, which may lead to the failure of detection of L dwarfs in these areas.
We noted that two undetected L dwarfs (spectral type L2, z =19.04134 mag and spectral type L4, z=19.04152 mag) were located in a dense region of the training sample. We examined these two L dwarfs and found their classification scores to be 0.68 and 0.77, respectively (In contrast, most of the undetected sources with scores close to 0), implying that they could be detected if the threshold (0.85) was lowered. The small number of L2 and L4 training samples  in this region may have led the model to give them insufficient attention, hence the scores for the two L dwarfs are slightly lower.


\section{Verification with more L dwarfs} 
\label{sec:Verification}
We further applied the LDAD model to a larger SDSS sample of L dwarfs which were selected according to the subclass in the SDSS spectral catalog. To obtain this sample, we selected sources with high galactic latitudes ($b$\textgreater45), with Class being "STAR" and subClass like "L" from the specObjAll and photoObjAll tables of SDSS DR16. A total of 2153 L dwarfs were obtained. Some samples with indistinct spectral characteristics of L dwarfs were removed by visually examining their spectra to ensure the purity of the sample. Then, L dwarfs without clean SDSS photometry or already in the training set and test set were removed from this sample set. Finally, 843 L dwarfs were left in 840 images.

\begin{figure}
\centering
    \includegraphics[width=0.45\textwidth]{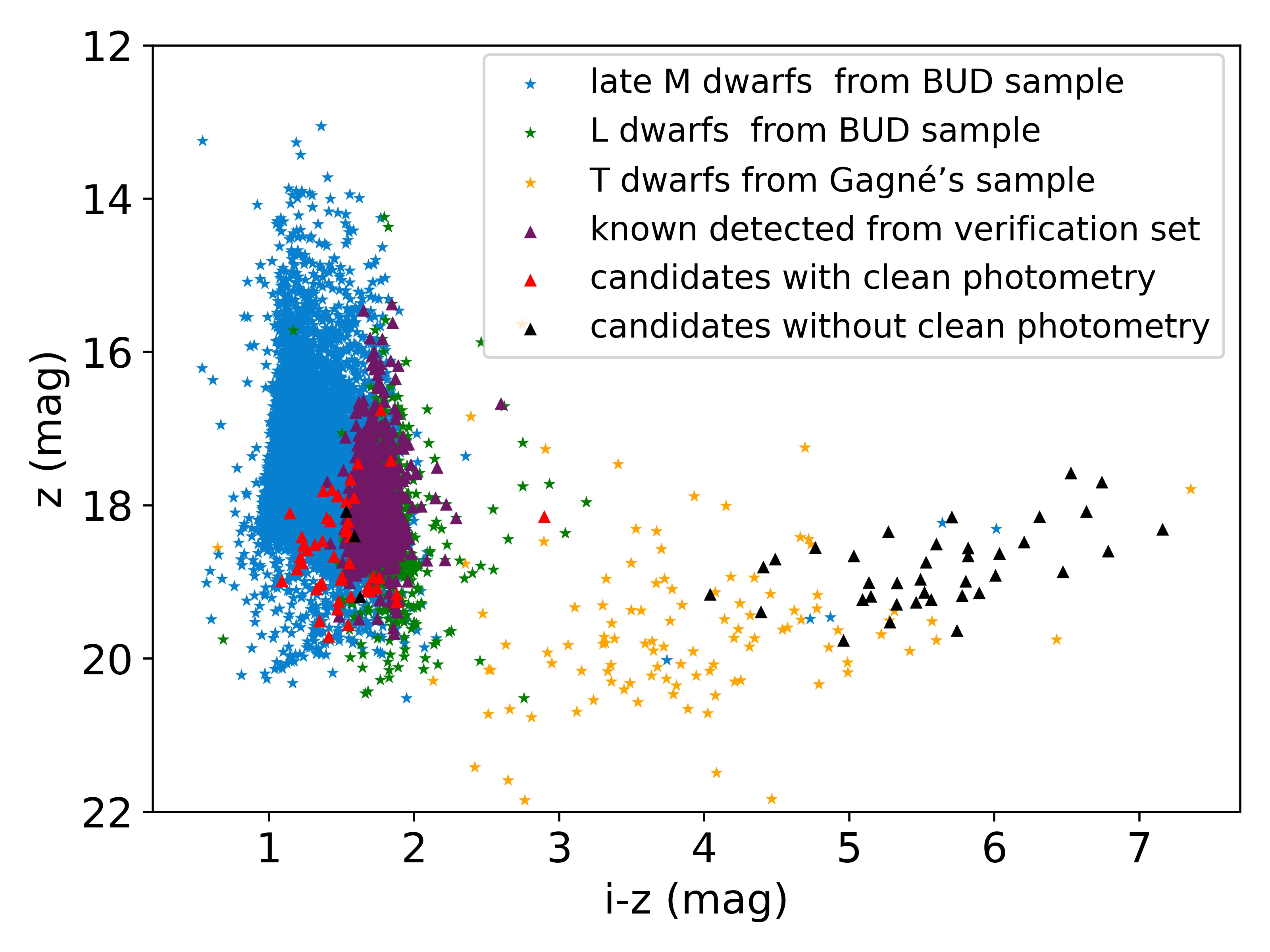}
\caption{Color-magnitude diagram of L dwarfs and L candidates detected by the LDAD model from the verification set.}
    \label{fig:i-z_ver}
\end{figure}

After preprocessing these images as previously described, the LDAD model was applied to detect L dwarfs from this sample.  As a result, 796 of 843 labeled L dwarfs were detected, and the recall rate was 94.42\%. In addition, 89 L dwarf candidates were detected. For these candidates, we estimated their spectral types using ($i-z$) colors, and the results are expressed in Table \ref{tab:validation result}. Among 89 candidates, there are 45 M dwarfs, 7 L dwarfs and 37 T dwarfs. The precision of the model for the validation set is 90.73\% while the contamination rate is 9.27\%.  Figure \ref{fig:i-z_ver} displays the color and magnitude distribution of L dwarfs and candidates detected by the LDAD model from the verification set. It can be seen from Figure \ref{fig:i-z_ver} that the detection results of the verification set are similar to those of the test set. The detected objects with clean photometry are mostly located in the L and late M color regions of the BUD sample.

For the validation set we obtained about the same model precision as the test set, however, the recall is higher. We attribute this to the better identifiability of L dwarfs in the validation set, as only L dwarfs with obvious spectral features were preserved through visual inspection of their SDSS spectra. (There are few faint L dwarfs in this dataset, and the incomplete coverage led us to abandon using it as a training dataset.) The validation experiment further demonstrates the effectiveness of the model in searching for L dwarfs from SDSS images.

\begin{table}\scriptsize \caption{LDAD detection results for
verification sample} \label{tab:validation result} \centering
\begin{tabular}{llllll} \hline

Class      & L & $M_{estimated}$& $L_{estimated}$  & $T_{estimated}$   \\
 \hline
Number     & 796     & 45                       & 7        & 37            \\
Proportion & 89.94\% & \textbf{5.09}\%                 & \textbf{0.79}\%    & \textbf{4.18}\%   \\
\hline
\end{tabular}
\end{table}

\section{CONCLUSIONS} 
\label{sec:CONCLUSIONS}
In this paper, we built the LDAD model using object detection techniques based on deep learning. Some popular techniques in computer vision such as deep convolution neural network, Resnet and FPN were utilized to develop
the LDAD model for detecting L dwarfs from SDSS field images. The LDAD model is data-driven, as it identifies L dwarfs and predicts their coordinates from SDSS images by learning features of 387 L dwarfs in SDSS images. The model adopts $riz$ images synthesized from $r, i$ and $z$ band images of SDSS as input, and outputs classification scores, central coordinates and bounding boxes of L dwarfs.

Applying the built LDAD model to the 93 SDSS images that contain 93 labeled L dwarfs in the test set, 83 known L dwarfs were detected successfully, with a recall rate of 89.25\%. For a verification set, in which the L dwarfs were chosen by the spectral types offered in the SDSS spectral catalog, 796 known L dwarfs were detected out of 843 labeled L dwarfs, with a recall rate of 94.42\%. In addition to recalling known L dwarfs, 10 and 89 L dwarf candidates were identified from the test set and the verification set respectively. Estimating their spectral type from color ($i-z$) reveals that they all belong to ultracool dwarfs, including late M, L and T dwarfs. 

The LDAD model has well learned the features of L dwarfs in SDSS images and can intelligently recognized L dwarfs from SDSS images without extracting photometric parameters in advance. In contrast to the common color cut method, the detection using the LDAD model avoids the errors introduced in the process of extracting photometric parameters. Thus, the detection using the LDAD model can complement the color cut results. The method proposed in this paper can be applied to upcoming imaging surveys to develop efficient object detection pipelines.

\section{ACKNOWLEDGMENTS} This study was supported by the Natural
Science Foundation of Shandong Province(Grant No. ZR2022MA089 and ZR2022MA076), the
National Natural Science Foundation of China (NSFC, Grant Nos. U1931209 and 11873037), and Cultivation Project for LAMOST
Scientific Payoff and Research Achievement of CAMS-CAS.

We thank SDSS for providing image and spectral data. Funding for the
SDSS has been provided by the Alfred P. Sloan Foundation, the
Participating Institutions, the National Science Foundation, the US
Department of Energy,  NASA, the Japanese Monbukagakusho, the Max
Planck Society, and the Higher Education Funding Council for
England. The SDSS Web Site is http://www.sdss.org.

\software{\texttt{pytorch} \citep{2019arXiv191201703P};
\texttt{matplotlib} \citep{Hunter:2007}; \texttt{astropy}
\citep{2013AA...558A..33A,2018AJ....156..123A} ; \texttt{reproject}
\citep{2018AJ....156..123A};  \texttt{numpy} \citep{van2011numpy}}


\bibliography{LDetect}{}
\bibliographystyle{aasjournal}



\end{document}